\documentclass[submission,copyright,creativecommons]{eptcs}
\providecommand{\event}{SCAV 2018} 
\usepackage{underscore}           
\usepackage{booktabs} 

\usepackage[ruled]{algorithm2e} 

\usepackage{amsmath,amssymb,amsfonts, amsthm}
\usepackage{algorithmic}
\usepackage{textcomp}
\usepackage{hyperref}
\usepackage{mathtools}
\usepackage{thmtools}

\usepackage{ae,aecompl}

\usepackage{calc}
\usepackage{epsfig}
\usepackage{graphicx}
\usepackage{framed}
\usepackage{stmaryrd}
\usepackage{xcolor}
\usepackage{bussproofs}
\usepackage{wrapfig}
\usepackage{enumitem}
\usepackage{listings}
\usepackage{tikz}
\usetikzlibrary{arrows,shapes,automata,petri,backgrounds,calc,intersections,decorations.pathreplacing}
\usepackage{wasysym}

 \usepackage{roadlogic}
 \usepackage{shortcuts}

\title{Introducing Liveness into Multi-lane Spatial Logic lane change controllers using UPPAAL\thanks{This research was partially supported by the German Research Foundation (DFG) in the Research Training Group GRK 1765 SCARE.}}

\author{Maike Schwammberger
\institute{Department of Computing Science, University of Oldenburg\\ Oldenburg, Germany\vspace{-0.2cm}}\\
\email{schwammberger@informatik.uni-oldenburg.de}
}

\def\titlerunning{Introducing Liveness into MLSL lane change controllers}
\def\authorrunning{M. Schwammberger}

\hypersetup{
	pdftitle={\@title},
	pdfauthor={\@author}
}

\begin{document}
\maketitle
\begin{abstract}
With Multi-lane Spatial Logic (MLSL) a powerful approach to formally reason about and prove safety of autonomous traffic manoeuvres was introduced. Extended timed automata controllers using MLSL were constructed to commit safe lane change manoeuvres on highways. However, the approach has only few implementation and verification results. We thus strenghen the MLSL approach by implementing their lane change controller in UPPAAL and confirming the safety of the lane change protocol. We also detect the unlive behaviour of the original controller and thus extend it to finally verify liveness of the new lane change controller.

\end{abstract}


\section{Introduction}\label{sec:introduction}
 Nowadays, driving assistance systems and fully autonomously driving cars are increasingly capturing the market. For such autonomous systems, traffic safety and prevention of human casualties is of the utmost importance. In this context, safety means collision freedom and thus reasoning about car dynamics and spatial properties. A softer, but also highly desirable, requirement is liveness, meaning that a good state is finally reachable.
 
 An approach to separate the car dynamics from the spatial considerations and thereby to simplify reasoning, was introduced in \cite{HLOR11} with the \MLSLL{} (\MLSL) for expressing spatial properties on multi-lane motorways with one driving direction for all cars. The idea to separate dynamics from control laws follows the work by Raisch et al. \cite{MRY02} and Van Schuppen et al. \cite{HCS06}.
 
The logic \MLSL{} and its dedicated abstract model was extended for country roads with oncoming traffic \cite{HLO13} and urban traffic scenarios with intersecting lanes \cite{HS16, Sch17}. The authors informally introduced respective controllers for safe lane change manoeuvres and safe turning manoeuvres at intersections. The respective safety of the controllers is proven with a semi-formal mathematical proof \cite{HLOR11, HLO13, HS16}. With \emph{automotive-controlling timed automata (ACTA)}, a formal semantics for the previously informal controllers was later introduced \cite{HS16}. (Un-) decidability results for (parts of) the logic \MLSL{} were provided \cite{FHO15, L15, O15}.
 
\MLSL{} itself is a thoroughly researched and strong formal approach for proving properties of autonomous traffic manoeuvres. Recently, the first computer-based assistance for reasoning with a new hybrid extension of \MLSL{} (HMLSL) was introduced by Linker \cite{L17}. The authors successfully investigate safety constraints for the motorway traffic scenarios from \cite{HLOR11} with Isabelle/HOL \cite{NPW03}. They outline an interesting extension of their work to liveness properties.

In this paper, we also focus on the motorway case. While \cite{L17} presents a strong implementation result focused directly on the spatio-temporal logic HMLSL, we instead investigate safety and liveness of the protocol of the lane change controller for highway traffic \cite{HLOR11}. The controller can be formalised as an automotive-controlling timed automaton (ACTA) \cite{HS16} and uses formulas of \MLSL{} to reason about traffic situations and to decide, whether a car can safely change lanes.

As ACTA are extended timed automata \cite{AD94}, we implement the lane change controller in the tool UPPAAL \cite{BDL04}, which allows for model-checking of timed automata. With this, we verify the correct behaviour of the considered lane change protocol and confirm the hitherto informally proven \emph{safety} property in a preferably generic UPPAAL model. Thus, our goal is to show \emph{unreachability} of a bad state with a collision in the overall system. With UPPAAL, we also detect the absence of \emph{liveness} in the original lane change controller from \cite{HLOR11}. We thus adapt the old lane change controller and show the liveness of the new controller with UPPAAL.

In Sect.~\ref{sec:preliminaries}, we briefly introduce the abstract model and logic \MLSL{} from \cite{HLOR11}. We also introduce the lane change controller and ACTA formalism. In Sect.~\ref{sec:uppaal}, we explain the adaptions of the lane change controller for the implementation in UPPAAL and introduce our UPPAAL verification properties. We extend the original controller from \cite{HLOR11} to a new live lane change controller in Sect.~\ref{sec:liveness}. Finally, we summarise our results in Sect.~\ref{sec:conclusion} and give ideas for future work.
\section{Preliminaries}\label{sec:preliminaries}
In this section, we briefly introduce the approach from \cite{HLOR11}. For this, we start with an overview over the abstract model for highway traffic in Sect.~\ref{sec:model} and introduce the \MLSLL{} in Sect.~\ref{sec:mlsl}. In Sect.~\ref{sec:acta}, we introduce the \etas{} (ACTA) from \cite{HS16}, which serve to formalise the lane-change controller from \cite{HLOR11}, that we describe in Sect.~\ref{sec:lane-change}.

\subsection{Abstract model and local view}\label{sec:model}
\begin{figure}[htbp]
\centering
	\includegraphics[scale=.95]{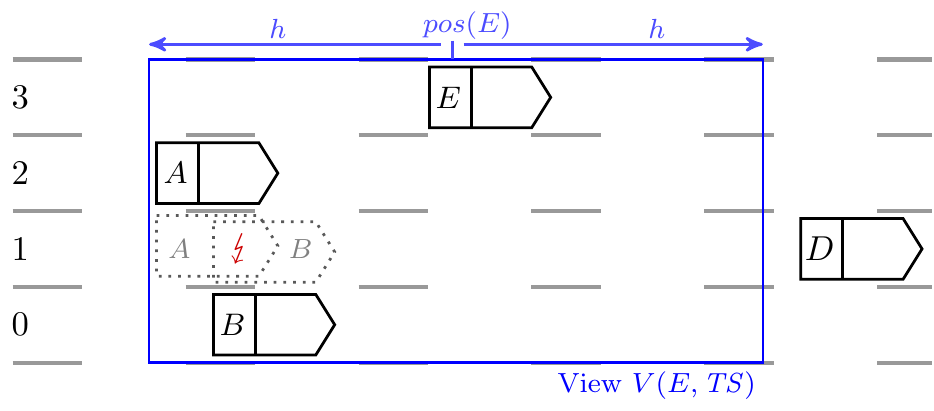}
	\caption[]{Abstract model with adjacent lanes $0$ to $3$ and cars $A$, $B$, $E$ and $D$. Cars $A$ and $B$ both plan to change to lane $1$, indicated with their resp. dotted claims on lane $1$. Car $D$ is too far away from car $E$ to be considered in $E$'s standard view $V(E, \mathit{TS})$.}
	\label{fig:model}
\end{figure}

The abstract model for highway traffic consists of neighbouring infinite \emph{lanes} $0, 1, \ldots$ of continuous space, leading in the same direction from the set of all lanes $\mathbb{L}$. Every car has a unique \emph{car identifier} $A, B, \ldots$ from the set $\mathbb{I}$ of all car identifiers and a real value for its position $\pos$ on a lane. An example for a traffic situation in our abstract model is depicted in Fig.~\ref{fig:model}. We use the concept of an \emph{ego car} as the \emph{car under consideration} and use the special variable $\ego$ to refer to this car. For Fig.~\ref{fig:model}, we assume $E$ is our ego car and thus have the valuation $\val(\ego) = E$.

In the abstract model, the space a car $E$ is currently occupying on a lane is represented by its \emph{reservation} $res(\ego)$, while a \emph{claim} $clm(\ego)$ is akin to setting the direction indicator (cf.\ dotted part of cars $A$ and $B$ in Fig.~\ref{fig:model}, showing the desire of $A$ and $B$ to change to lane $1$). Thus, a claim represents the space a car plans to drive on in the future. For now we assume, that the size of a car includes its' physical size and its braking distance. With this, safety is already violated, if a car invades the braking distance of another car. The idea is that every car is supposed to be able to do an emergency brake at every moment, without causing a collision.

Static information about cars like their positions and their reserved or claimed lanes is captured in a \emph{\traffics{}} $\RoadDef$ from the set $\mathbb{TS}$ of all traffic snapshots. E.g. $res(A)= \{2\}$, $clm(A)=\{1\}$ and $pos(A)=10$ for car $A$ in Fig.~\ref{fig:model}. As lanes are of infinite size, we also have an infinitely large traffic snapshot with infinitely many cars in it. However, for checking safety and liveness properties of our lane-change controller, only cars within some bounded \emph{view} $V$ around our ego car $E$ are of interest.
\begin{definition}[View]\label{def:view}
For an arbitrary \traffics{} $\Road$, the \emph{view} $V$, owned by car $E\in\mathbb{I}$, is defined by $V=(\ViewLanes,X,E)$, where $\ViewLanes \subseteq \mathbb{L}$ is an interval of lanes visible in $V$ and $X= [r,t]\subseteq \R$ is an interval of space along the lanes.

We define the \emph{standard view} of car $E$ by $V (E, \Road) =(\mathbb{L}, [\pos(E) - h , \pos(E) + h] ,E)$, where $h$ is a sufficiently large \emph{horizon} for looking forwards resp.\ backwards from the position $\pos(E)$, as given in the traffic snapshot \Road.
\end{definition}
Note that we assume there exists a minimal positive value for the size of all cars, thus only finitely many cars are considered in a view. We furthermore assume that there exists a maximum velocity for all cars and the horizon $h$ is big enough to consider the fastest car that could endanger $E$ contained in its the standard view $V(E,\Road)$. In the example in Fig.~\ref{fig:model}, car $D$ is not considered in $V$, as it is to far away from $E$.

We use a car dependent sensor function $\Omega_E \colon \ID \times \mathbb{TS} \rightarrow \R_+$ which, given a car identifier $C\in\ID$ and a traffic snapshot $\Road\in\mathbb{TS}$, provides the size $\Omega_E (C, \Road)$ of $C$ as perceived by $E$'s sensors.

For a view $V=(\ViewLanes,X,E)$ and a traffic snapshot \RoadDef, we introduce the following abbreviations, used for the semantics definition of our logic \MLSL{} in the next Sect.~\ref{sec:mlsl}:
\begin{align}
    {res}_V \colon\ID\rightarrow\mathbb{P}(\ViewLanes) &\text{ with } {res}_V (C) = res(C) \cap L\label{formula:resv}\\
    {clm}_V \colon\ID\rightarrow\mathbb{P}(\ViewLanes) &\text{ with } {clm}_V (C) = clm(C) \cap L\label{formula:clmv}\\
    {len}_V \colon\ID\rightarrow\mathbb{P}(\ViewLanes) &\text{ with } {len}_V (C) = [pos(C), pos(C)+{\Omega}_E (C,\Road)] \cap X\label{formula:lenv}
\end{align}
The functions \eqref{formula:resv} and \eqref{formula:clmv} restrict their counterparts $res(C)$ and $clm(C)$ from \Road{} to the set of lanes considered in $V$. Function \eqref{formula:lenv} defines the part of car $C$ that $E$ perceives with its sensors in the extension $\Extension$ of the considered view $V$.

\subsection{\MLSLL}\label{sec:mlsl}
With \MLSLL{} (\MLSL), we can reason about traffic situations in our local view $V$. As variables, we allow for \emph{car variables} $c,d , \ldots$ from the set $\carvariables$, valuated with car identifiers from the set $\mathbb{I}$ and \emph{lane variables} $n,l, \ldots$ from the set $\lanevariables$, valuated with lanes from $\mathbb{L}$. We define $\ego \in \carvariables$.
\begin{definition}[Valuation of variables]
A \emph{valuation} $\nu$ is a function
\(\val \colon \variables \to \ID\cup\mathbb{L} \), where $\variables = \carvariables\cup\lanevariables$ and \(\val \colon \carvariables \to \ID \) and \(\val \colon \lanevariables \to \mathbb{L} \).
\end{definition}

Formulae of \MLSL{} are built from atoms, Boolean connectors and first-order quantifiers. As \emph{spatial} atoms, we use $\free$ to represent free space on a lane and $re(c)$ (resp. $cl(c)$) to formalise the reservation (resp. claim) of a car. We also allow for the comparison of variables $u=v$ for variables $u,v, \in\variables$ of the same type.

We use a horizontal \emph{chop operator} similar to chop operations for timing intervals in Duration Calculus \cite{ZHR91} or interval temporal logic \cite{Mos85}, denoted by $\chop$. Also, we introduce a vertical chop operator given by the vertical arrangement of formulas. Intuitively, a formula $\varphi_1 \chop \varphi_2$ holds if we can split the view $V$ vertically into two views $V_1$ and $V_2$ such that on $V_1$ the formula $\varphi_1$ holds and $V_2$ satisfies $\varphi_2$. Similarly a formula ${}_{\varphi_1}^{\varphi_2}$ is satisfied by $V$, if the view can be chopped at a lane into two subviews, $V_1$ and $V_2$, where $V_i$ satisfies $\varphi_i$ for $i=1,2$.

\begin{definition}[Syntax]\label{def:mlsl}
The syntax of a \emph{\MLSLL} formula ${\phi}_M$ is defined by
  \begin{align*}
    {\varphi}_M ::= &\true\mid u=v \mid\free\mid\reserved{c} \mid\claimed{c} \mid\lnot\varphi\mid\varphi_1 \land\varphi_2
	      \mid\exists c \colon\varphi_1 \mid\varphi_1 \chop\varphi_2 \mid {}_{\varphi_1}^{\varphi_2}\text{,}
  \end{align*}
where $c \in \carvariables$ and $u,v \in \variables$. We denote the set of all \MLSL{} formulas by \({\formulae}_{\mathbb{M}}\).
\end{definition}
The semantics of \MLSL{} formulas is defined over a \traffics{} \Road, a view $V$ and a valuation of variables $\val$. We denote the length of a real interval $X\subseteq\R$ by $|X|$.

\begin{definition}[Semantics of \MLSL]\label{def:semantics-MLSL}
The \emph{satisfaction} of \MLSL{} formulas $\varphi$ with respect to a traffic snapshot $\Road$, a view $V = (\ViewLanes,\Extension,E)$ with $\ViewLanes = [l,n]$ and $\Extension = [r,t ]$, and a valuation $\val$ of variables is defined inductively as follows:
\allowdisplaybreaks
\begin{align*}
\Road,V,\val&\models \true & & \:\text{ for all } \Road, V, \val\\
\Road,V,\val&\models u = v & \Leftrightarrow &\: \val(u) = \val(v)\\
\Road,V,\val&\models \free & \Leftrightarrow &\: |\ViewLanes| =1 \text{ and } |\Extension| > 0 \text{ and } \forall i\in I_V \colon len_V (i) \cap (r,t) = \emptyset \\
\Road,V,\val&\models \reserved{c} & \Leftrightarrow &\: |\ViewLanes| =1 \text{ and } |\Extension| > 0 \text{ and } \val (c) \in I_V \text{ and }\, {res}_V (\val(c)) = \ViewLanes \text{ and } \Extension = {len}_V (\nu (c)) \\
\Road,V,\val&\models \claimed{c} & \Leftrightarrow & \: |\ViewLanes| =1 \text{ and } |\Extension| > 0 \text{ and } \val (c) \in I_V \text{ and }\, {clm}_V (\val(c)) = \ViewLanes \text{ and } \Extension = {len}_V (\nu (c)) \\
\Road,V,\val&\models  \neg\varphi & \Leftrightarrow &\: \text{not } \Road,V,\val\models \varphi \\
\Road,V,\val&\models  \varphi_1 \land \varphi_2  & \Leftrightarrow &\: \Road,V,\val\models\varphi_1 \text{ and } \Road,V,\val\models\varphi_2 \\
\Road,V,\val&\models  \exists\colon\varphi_1 & \Leftrightarrow &\: \Road,V,\val\models \exists\alpha \in I_V \colon \Road,V,\val\oplus \{c \mapsto \alpha\} \models\varphi_1 \\
\Road,V,\val&\models \varphi_1\chop\varphi_2&\Leftrightarrow &\: \exists s\in\R \colon r \leq s \leq t \text{ and }\Road,V_{[r,s]}, \val \models \varphi_1\text{ and } \Road,V_{[s,t]},\val\models \varphi_2\\
\Road,V, \val&\models \text{\footnotesize{$\begin{array}{c}{\varphi_2}\\{\varphi_1}\end{array}$}}&\Leftrightarrow & \: \exists m \in\N\colon l-1 \leq m \leq n+1 \text{ and } \Road,V^{[l,m]}, \val \models \varphi_1\text{ and } \Road,V^{[m+1,n]},\val\models \varphi_2\\
\end{align*}
\end{definition}

\textit{Abbreviation.} In the following we use the abbreviation $\langle \varphi \rangle$ to state that a formula $\varphi$ holds \emph{somewhere} in the considered view. For example, in Fig.~\ref{fig:model} with valuation $\nu(\ego)=E$, the formula $\langle\varphi\rangle \equiv \langle \reserved{\ego}\rangle$ holds in $V(E,Road)$, because there \emph{somewhere} exists a reserved space for car $E$. 

\begin{example}[\MLSL{} formulas]\label{ex:mlsl}
Consider Fig.~\ref{fig:model} and assume a valuation of variables $\nu(\ego) = E$, $\nu(a) = A$, $\nu(b) =B$ and $\nu(d)=D$. Consider the following \MLSL{} formulas:
 \begin{align*}
    {\varphi}_1 &\equiv \langle \reserved{\ego} \chop\free \rangle \\
    {\varphi}_2 &\equiv \langle\claimed{a} \land \claimed{b} \chop \neg\claimed{a}\land\claimed{b}\rangle \\
    {\varphi}_3 &\equiv \langle\claimed{b} \chop\free\chop\reserved{d} \rangle
 \end{align*}
In view $V(E,\Road)$ the formula ${\varphi}_1$ holds, as there is free space in front of car $E$. Equally ${\varphi}_2$ holds, as there is a claim of both cars $A$ and $B$ at the same spot on lane $1$ and after this there is a space with only the claim of car $B$. Thus $\Road, V(E,\Road), \nu\models {\varphi}_1$ and $\Road, V(E,\Road), \nu\models {\varphi}_2$. However, $\Road, V(E, \Road), \nu\not\models {\varphi}_3$, as car $D$ is not part of view $V(E,\Road)$.
\end{example}

\subsection{Automotive-controlling timed automata}\label{sec:acta}
Before we introduce the actual lane change controller protocol from \cite{HLOR11} in Sect.~\ref{sec:lane-change}, we briefly define the extended timed automata type, introduced in \cite{HS16} to formalise the controller. As variables these \etas{} (ACTA) use both clock and data variables. For clock variables $x,y \in \mathbb{X}$ and clock updates we refer to the definition of timed automata \cite{AD94} and for data variables $u,v \in\variables$ and data updates we refer to the extension of timed automata proposed for UPPAAL \cite{D15}. These clock and data updates ${\nu}_{act}$ are allowed on transitions of ACTA.

Further on, the controllers use \MLSL{} formulas ${\varphi}_{M}$ as well as clock and data constraints ${\varphi}_\mathbb{X}$ resp. ${\varphi}_{Var}$ as guards $\varphi$ on transitions and invariants $I(q)$ in states $q$. An example for a data constraint for a variable $l \in Var$ is $l > 1$. A guard or invariant $\varphi$ from the set $\Phi$ of all guards and invariants is defined by $\varphi \:\equiv\; {\varphi}_{M} \;|\; {\varphi}_\mathbb{X} \;|\; {\varphi}_{Var} \;|\; {\varphi}_1 \wedge {\varphi}_2 \;|\; true\text{.}$

We express possible driving manoeuvres by \emph{controller actions}, which may occur at the transitions of an \acta. Controller actions e.g.\ enable a car to set or withdraw ($\mathtt{wd}$) a claim ($\mathtt{c}$) or a reservation ($\mathtt{r}$) for a lane.
\begin{definition}[Controller Actions]\label{def:controlleractions}
	With $c \in\carvariables$, a \emph{controller action} $c_{act}$ is defined by
	\begin{align*}
		\begin{array}{c}
		      c_{act} ::= \mathtt{c}(c, \datapsi) \mid \mathtt{wd} \; \mathtt{c}(c) \mid \mathtt{r}(c) \mid \mathtt{wd} \; \mathtt{r}(c, \datapsi) \mid \tau \text{,}
		\end{array}
	\end{align*}
	where $\datapsi ::= k \; | \; l_1 \; | \; l_1 + l_2 \; | \; l_1 - l_2$ with $k \in \mathbb{N}$, $l_1 , l_2 \in\lanevariables$. The set of all controller actions is defined by ${Ctrl}_{Act}$.
\end{definition}

\subsection{Lane change controller}\label{sec:lane-change}
In this section, we introduce the lane change controller from \cite{HLOR11}, whose implementation into UPPAAL we introduce in Sect.~\ref{sec:uppaal}. The overall goal of this controller is to safely change lanes in freeway traffic. Here, \emph{safety} of ego car means collision freedom and thus disjunction of the reserved spaces of ego and other cars, expressed by the \MLSL{} formula
\begin{align}\label{formula:safe}
\text{\emph{Safe}}(ego) \;\equiv\; \neg \exists c \colon c \neq ego \land \somewhere{\reserved{ego} \land \reserved{c}}\text{.}
\end{align}

The main idea for the lane change controller is to first \emph{claim} the space on a lane it wants to enter and \emph{reserve} it only if no collision is detected. We assume a lane change to take at most $t_{lc}$ time to finish. The lane change controller is constructed for the ego car ($\nu(\ego) = E$ in the example from Fig.~\ref{fig:model}) but scales to all cars as $\ego$ can be substituted by an arbitrary car variable $c \in\carvariables$.

We explain the construction of the controller starting with the initial state. As we want to prevent different reservations from overlapping, we introduce a \emph{collision check} for the ego car expressed by the \MLSL{} formula
    \begin{align}\label{formula:cc}
	cc \; \equiv \; \neg\exists c \colon c \neq \ego \wedge \langle re(\ego) \wedge re(c) \rangle \text{.}
    \end{align}
Formula \eqref{formula:cc} is evaluated to true, iff nowhere exists a car different from the ego car whose reservation overlaps with the actors reservation. We assume $cc$ to hold in the initial state of our controller. Next the lane change controller can claim some space on either the lane to its left or right, provided such a lane exists. Here $N$ is the lane identifier of the highest lane from the set of all lanes $\mathbb{L}$.

In order to transform a claim into a reservation and thus finally change lanes, a car first needs to check if there are overlaps of other cars' claims or reservations with its own claim. This is formalised by the \emph{potential collision check}
    \begin{align}\label{formula:pc}
	pc(c) \; \equiv \; c \neq \ego \wedge \langle cl(\ego) \wedge (re(c) \vee cl(c)) \rangle \text{.}
    \end{align}
Formula \eqref{formula:pc} evaluates to true, iff there exists a car different from the ego car whose claim or reservation overlaps with ego car's own claim. A (temporary) potential collision is allowed, because it does not endanger the safety property \eqref{formula:safe}. However, if a potential collision is detected, the car must withdraw its claim immediately.

When $\exists c \colon pc(c)$ does not hold, the actor reserves the claimed lane and starts changing lanes. To prevent deadlocks, we set a time bound $t$ in state $q_2$ for the time that may pass between claiming and reserving crossing segments. After $t_{lc}$ time, the lane change is finished and the reservation of actor $E$ is reduced to the new lane.
\begin{figure}[htbp]
\vspace{-0.0cm}
\centering
	\begin{tikzpicture}[initial text=,->,>=stealth',shorten >=1pt,auto,node distance=3.5 cm,
                semithick, scale = 1.0, transform shape,inner sep=0.8pt,minimum size=20pt,]
        \tikzstyle{state}=[draw, shape=ellipse]
	\begin{scope}
		\node [state, initial]	(q0) [label=above left:]{$q_0 : cc$};
		\node [state]		(q1) [xshift=+1.0cm, right of=q0, label=above right:]{$q_1$};
		\node [state]		(q2) [xshift=+1.0cm, right of=q1, label=above right:]{$q_2 :$ \parbox{1.76cm} {\centering{$\neg \exists c: pc(c)$} \\ \centering{$\wedge x \leq t$}}};
		\node [state]		(q3) [xshift=-0.0cm, yshift=0.2cm, below of=q2, label=above right:]{$q_3: x \leq t_{lc}$};

	    	\path	(q0) edge [bend left=00]  node[below, xshift=+0.0cm, yshift=-0.05cm] {\parbox{3cm}{\centering{$n+1 \leq N$} \\ \centering{$/$ \texttt{c}$(ego, n+1);$} \\ \centering{$l := n+1$}}} (q1)
			(q0) edge [bend right=10, min distance=22mm,out=-75,in=-90]  node[right, xshift=+0.8cm, yshift=-0.1cm] {\parbox{3cm}{\centering{$0 \leq n-1$} \\ \centering{$/$ \texttt{c}$(ego, n-1);$} \\ \centering{$l := n-1$}}} (q1)
	    		(q1) edge [bend right=30]  node[above, yshift=-0.1cm] {\parbox{3cm}{\centering{$\exists c\colon pc(c)$}\\\centering{\texttt{wd c}$(\ego)$}}} (q0)
	    		(q1) edge [bend left=00]  node[above] {\parbox{3cm}{\centering{$\neg\exists c\colon pc(c)$} \\\centering{$/x:=0$}}} (q2)
 	    		(q2) edge [bend right=20, looseness=0.65, out=-80,in=-90]  node[above, xshift=+0.0cm, yshift=-0.2cm] {$\exists c : pc(c) /$\texttt{wd c}$(\ego)$} (q0)
 	    		(q2) edge [bend left=00]  node[right, xshift=0.1cm, yshift=-0.00cm] {\parbox{3cm}{$\neg\exists c : pc(c)$\\$/$\texttt{r}$(\ego);$\\$x:=0$}} (q3)
 	    		(q3) edge [bend left=65, looseness=1.1, min distance=25mm,out=30,in=108]  node[left, xshift=-0.2cm, yshift=+0.35cm] {\parbox{3cm}{$x \geq t_{lc}/$ \\ \texttt{wd r} $(\ego);$ \\ $n:=l$}} (q0);
	\end{scope}
	\end{tikzpicture}
	\vspace{-0.0cm}
	\caption[]{Lane change controller from \cite{HLOR11}.}
	\label{fig:lcp}
\end{figure}
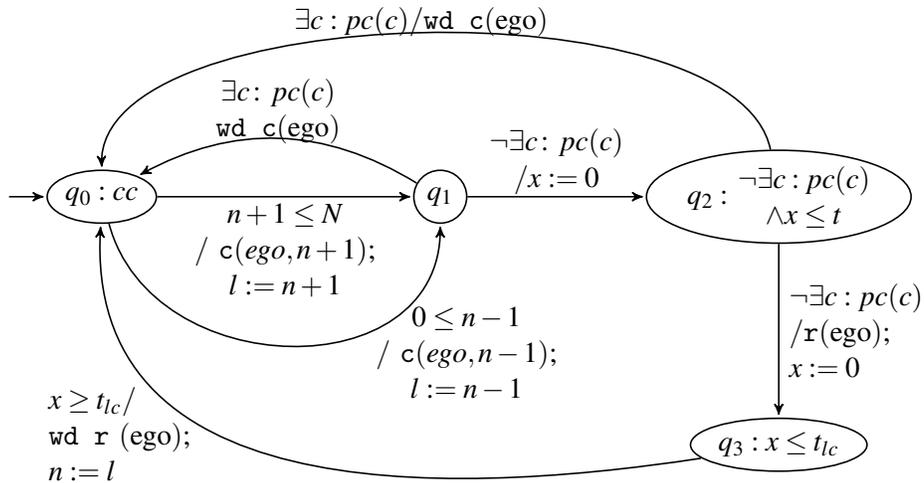
\section{UPPAAL Implementation and Verification}\label{sec:uppaal}
We first introduce the specific abstract model we examine with UPPAAL and the considered assumptions and restrictions for it in Sect.~\ref{sec:uppaal-model}. We explain the adaptions of the lane change controller from Sect.~\ref{sec:lane-change} to the type of extended timed automata UPPAAL accepts in Sect.~\ref{sec:implementation}. We explain our verification method and show safety of the existing controller in Sect.~\ref{sec:verification}. We detect liveness issues for the lane change controller from Sect.~\ref{sec:lane-change} and adapt it to a live controller in Sect.~\ref{sec:liveness}. We provide a summary of the goals and limitations of the current implementation and give an overview over scenario and UPPAAL model extensions in Sect.~\ref{sec:extendability}.

\subsection{UPPAAL-Model and Assumptions}\label{sec:uppaal-model}
\subsubsection{Overall scenario and data structure}
The model we examine with UPPAAL is the traffic situation depicted in Fig.~\ref{fig:model}, where we consider lanes $0$ to $3$ and the cars $A$, $B$ and $E$ contained in view $V(E,\Road)$. We encode the \traffics{} \Road, more precisely the positions, claims and reservations of the cars on the lanes, by a global data structure $pos\_t$. For reservations $res$ this is encoded as follows:
\begin{lstlisting}
pos_t res[carid_t] = {
		      { {0,0,1,0}, 10, 5}, 
		      { {1,0,0,0}, 12, 5}, 
		      { {0,0,0,1}, 40, 5}
		     };
\end{lstlisting}
Here e.g.\ the first line represents car $A$ and the Boolean lane list $\{0,0,1,0\}$ states that $A$ has a reservation only on lane $2$. The second parameter $10$ is the position of $A$ on lane $2$ and the last parameter $5$ is the size of $A$. Thus the space $A$ occupies is the interval $[10,15]$ on lane $2$. The other lines are the respective values for cars $B$ and $E$, such that $B$ initially occupies interval $[12,17]$ on lane $0$ and $E$ occupies interval $[40,45]$ on lane $3$. We have a similar structure $\mathtt{pos\_t \; clm[carid\_t]}$ for the claims of the cars, where initially all Boolean lists for claims are empty, as all cars are supposed to start in the initial state of the controller without any claim.

\subsubsection{Distance Controller} 
\begin{figure}[htbp]
\centering
	\includegraphics[scale=1.0]{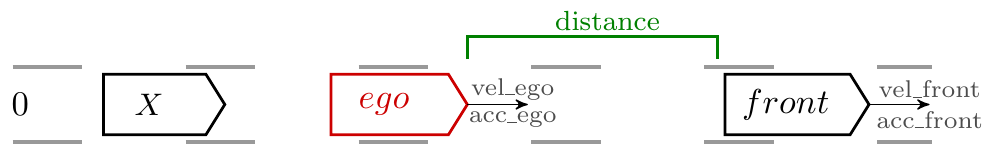}
	\caption[]{One-lane scenario with distance keeping from \cite{LMH15}.}
	\label{fig:upp:distance}
\end{figure}

The lane change controller is not responsible for distance keeping. However, for cars with different acceleration and speed, a controller for distance keeping is inevitable to avoid rear-end collisions. Such a distance controller is outlined, but not formalised or constructed in \cite{HLOR11}. Another possible distance controller is introduced and formally verified, but not yet implemented in \cite{DHO06}. Recently, the group of Kim Larsen synthesised an adaptive cruise control distance controller with the UPPAAL extension Stratego \cite{LMH15}. As the authors base their work on the spatial model of \MLSL, this approach is of high interest for our implementation. However, they only consider a model consisting of one single lane without any neighbouring lanes and only two specific cars $ego$ and $front$ (cf.~Fig.~\ref{fig:upp:distance}). Their idea is, that the $ego$ car keeps track of its distance to the $front$ car \emph{always}. Additionally, their goal is to minimise the distance between $ego$ and $front$. For this, one UPPAAL automaton for each $ego$ and $front$ is used, additional to a system controller. 

\begin{figure}[htbp]
\centering
	\includegraphics[scale=.95]{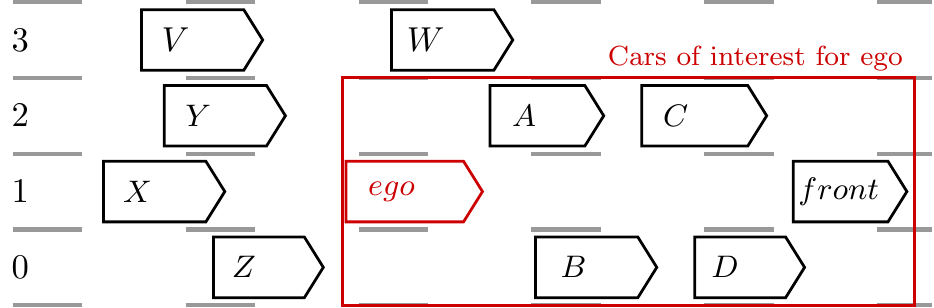}
	\caption[]{Cars of interest for ego car for distance keeping in multi-lane highway scenario \cite{HLOR11}.}
	\label{fig:model2}
\end{figure}

Consider on the other hand our multi-lane scenario, e.g. in Fig.~\ref{fig:model2}. It is not enough to keep track of the distance to $front$, as cars $A$, $B$, $C$ and $D$ might change lanes and thus be in front of $ego$ any time. Thus, we also need to keep track of the distances to these cars. A problem here is state space explosion, as the number of considered parallel timed automata for UPPAAL increases significantly, when using the approach from \cite{LMH15} directly. A second problem is the discretisation of space in their approach.

However, for examining the safety and liveness solely of the lane change manoeuvres with the controller from \cite{HLOR11}, we do not need to consider a scenario with cars with different speed and acceleration. We restrict all cars to have the same constant speed whereby the relative distances between the cars along the lanes never change. Although this is a strong restriction, it is reasonable, as our goal is to show safety and liveness of lane change manoeuvres, where collision freedom while changing lanes is considered, not rear-end collisions.

Nonetheless, as a constant speed for all cars is a strong assumption, we plan to implement a version of the adaptive cruise controller from \cite{LMH15} in future work for a more realistic model.

\subsubsection{Generic model} Despite the speed limitation, we encode a preferably general behaviour. In our model, the expected behaviour of car $E$ is that it is \emph{always} able to change lanes and that there can \emph{never} occur a \emph{potential collision or collision} with $E$, as there is no conflicting car on any neighbouring lane. In contrast, cars $A$ and $B$ can \emph{not always} change lanes, as their position intervals $[10,15]$ and $[12,17]$ would intersect if the cars had reservations or claims on the same lane. Thus, we \emph{expect potential collisions} between $A$ and $B$, but show that the lane change controller \emph{always} prevents \emph{actual collisions}.

\subsection{Implementation}\label{sec:implementation}
\begin{figure}[htbp]
\centering
	\includegraphics[scale=.65]{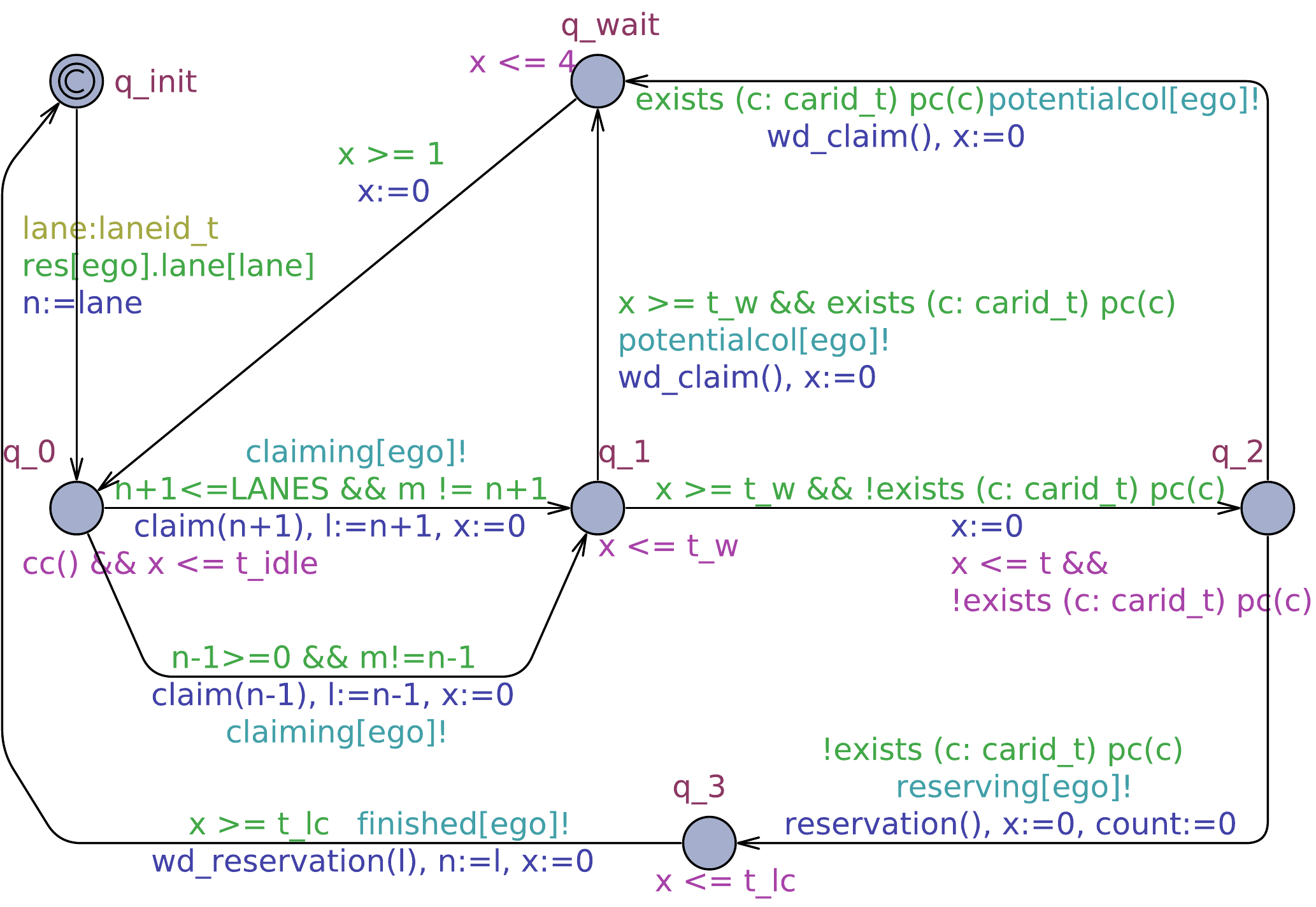}
	\caption[]{Lane-change controller implementation \texttt{LCP} in UPPAAL}
	\label{fig:upp:LCP}
\end{figure}

For the UPPAAL implementation, we adapt the lane change controller from Fig.~\ref{fig:lcp} to UPPAAL syntax, as neither formulas of \MLSLL{} (cf.\ Def.~\ref{def:mlsl}, p.~\pageref{def:mlsl}) nor controller actions for claiming or reserving lanes (cf.\ Def.~\ref{def:controlleractions}, p.~\pageref{def:controlleractions}) are directly implementable in UPPAAL. The resulting UPPAAL lane change controller \texttt{LCP} is depicted in Fig.~\ref{fig:upp:LCP}. Each of the cars $A$, $B$ and $E$ in our model owns one instance \texttt{LCP(i)} of the controller \texttt{LCP}, where \texttt{i} ranges over $A$, $B$ and $E$. Note, that Fig.~\ref{fig:upp:LCP} already contains the adaptions to a live controller, we explain later in Sect.~\ref{sec:liveness}.

\vspace{-0.0cm}
We start with the UPPAAL representation of MLSL formulas. The only MLSL formulas used by the lane change controller are the collision check $cc$ (cf.\ formula \eqref{formula:cc}, p.~\pageref{formula:cc}) in the initial state $q_0$ and the potential collision check $pc(c)$ (cf.\ formula \eqref{formula:pc}, p.~\pageref{formula:pc}) used in several guards and invariants of the controller. Our solution for implementing formulas \eqref{formula:cc} and \eqref{formula:pc} in UPPAAL bases on checking the intersection of position intervals of cars with the Boolean UPPAAL function
\begin{lstlisting}
bool intersect(const pos_t p1, const pos_t p2) {
   return exists(lane: laneid_t) 
      p1.lane[lane] and p2.lane[lane] 
        and not (p1.pos > p2.pos+p2.size or p2.pos > p1.pos+p1.size);
}
\end{lstlisting}
\noindent
The function \texttt{intersect} checks for two position parameters \texttt{pos\_t} (cf.\ Sect.~\ref{sec:uppaal-model}) if their position intervals intersect and if both positions are on the same lane. If e.g.\ car $A$ and $B$ both claim lane $1$ with $\mathtt{clm[A]=\{ \{0,1,0,0\}, 10, 5\}}$ and $\mathtt{clm[B]= \{ \{0,1,0,0\}, 12, 5 \}}$, the function call  $\mathtt{intersect(clm[A], clm[B])}$ returns $\mathtt{true}$.
\par\medskip

With the intersect function, we encode the collision check $cc$ from MLSL formula \eqref{formula:cc} by the function
\begin{lstlisting}
bool cc () {
   return not exists(c:carid_t) c != ego
	  and intersect(res[ego],res[c]);
}
\end{lstlisting}
\noindent
and the potential collision check $pc(c)$ from MLSL formula \eqref{formula:pc} with
\begin{lstlisting}
bool pc (carid_t c) {
   return c != ego
	  and (intersect(clm[ego],res[c])
	       or intersect(clm[ego],clm[c]));
}
\end{lstlisting}
\noindent
We use the functions $\mathtt{cc()}$ and $\mathtt{pc(c)}$ in the UPPAAL controller \texttt{LCP} in Fig.~\ref{fig:upp:LCP} exactly in the same manner as we use the respective MLSL formulas in the original lane change controller from Fig.~\ref{fig:lcp}. Besides MLSL formulas, we also encode controller actions for claiming and reserving lanes and their respective withdrawal actions with UPPAAL methods. For claiming a lane for the ego car, the related lane change controller calls the method
\begin{lstlisting}
 void claim(laneid_t lane) {
    clm[ego].lane[lane] = true;
}
\end{lstlisting}
\noindent
where in the Boolean list $\{0,0,0,0\}$ for claims, the value of the forwarded lane $\mathtt{lane}$ is set to $\mathtt{true}$. Upon a reservation request from a lane change controller, we have to check if there exists a claim for the related car and only then transform the claim into a reservation. Thus,
\begin{lstlisting}
void reservation(){
    for (i:laneid_t)
    {
        if (clm[ego].lane[i]) {
            res[ego].lane[i] = true;
            clm[ego].lane[i] = false;
        }
    }
}
\end{lstlisting}
\noindent
changes the value of the respective lane in the reserved lanes for the ego car to $\mathtt{true}$, while setting the value for the transformed claim for the same lane to $\mathtt{false}$.
\subsection{Verification of Safety with UPPAAL}\label{sec:verification}
The requirement queries for the verifier in UPPAAL are formulated in a computation tree logic (CTL) \cite{CE82, QS82} style specification language. The first query we successfully check is
\begin{align}\label{q1}
  \text{\texttt{A[] not deadlock},}
\end{align}
with which we globally exclude deadlocks in an arbitrary run of our system. We checked the query on a normal work station in $48$ to $49$ seconds with a memory usage peak of roughly $140$KB.

\begin{figure}[htbp]
\centering
	\includegraphics[scale=.4, trim=0 0.6cm 0 0.4cm]{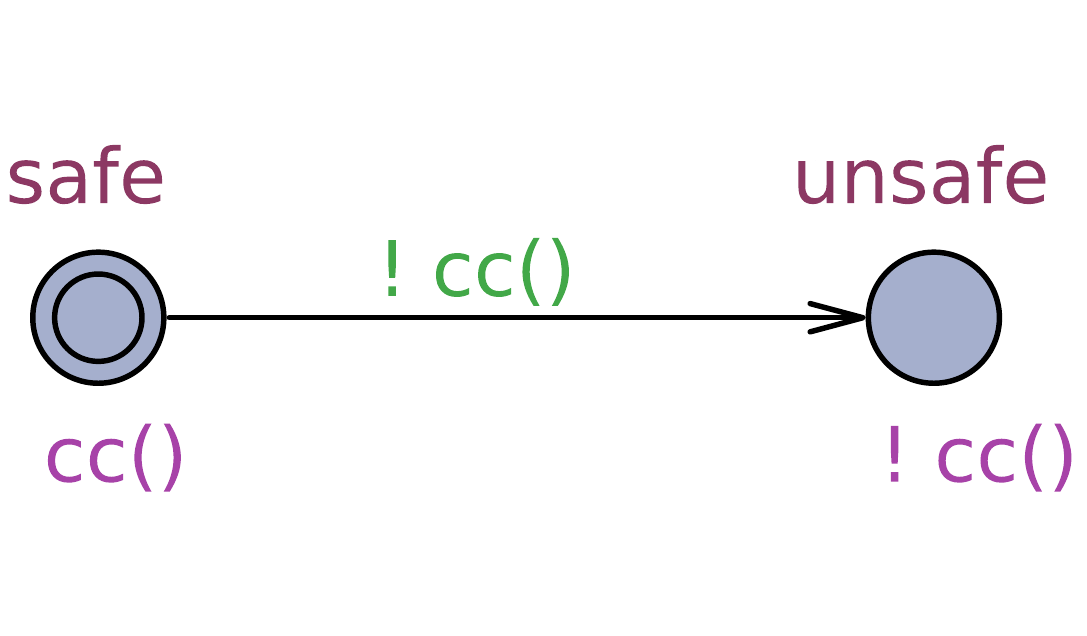}
	\caption{\texttt{Observer1} checking for a collision.}
	\label{fig:upp:Observer2}
\end{figure}

For the second query, we introduce the Observer automaton \texttt{Observer1}, depicted in Fig.~\ref{fig:upp:Observer2}. This Observer automaton uses a slightly adapted version of the collision check \texttt{cc()} to check for a collision between any two arbitrary cars at any moment. We use the query
\begin{align}\label{q2}
  \text{\texttt{A[] not Observer1.unsafe}}
\end{align}
to show in averagely less than $4$ seconds with a memory usage peak of $46$KB, that there exists no example trace where the formula $cc$ does not hold. With this query, we verify the safety property \eqref{formula:safe} (p.~\pageref{formula:safe}) for the lane change controller from \cite{HLOR11}.

\subsection{Adaptions for constructing a live controller}\label{sec:liveness}
With query \eqref{q1}, we exclude deadlocks in our system. However, the original controller in Fig.~\ref{fig:model} is not truly live, as e.g.\ \emph{livelocks} exist, where no car ever changes lanes, even though in our model at least car $E$, should always be able to change lanes.

\begin{figure}[htbp]
\centering
	\includegraphics[scale=.65]{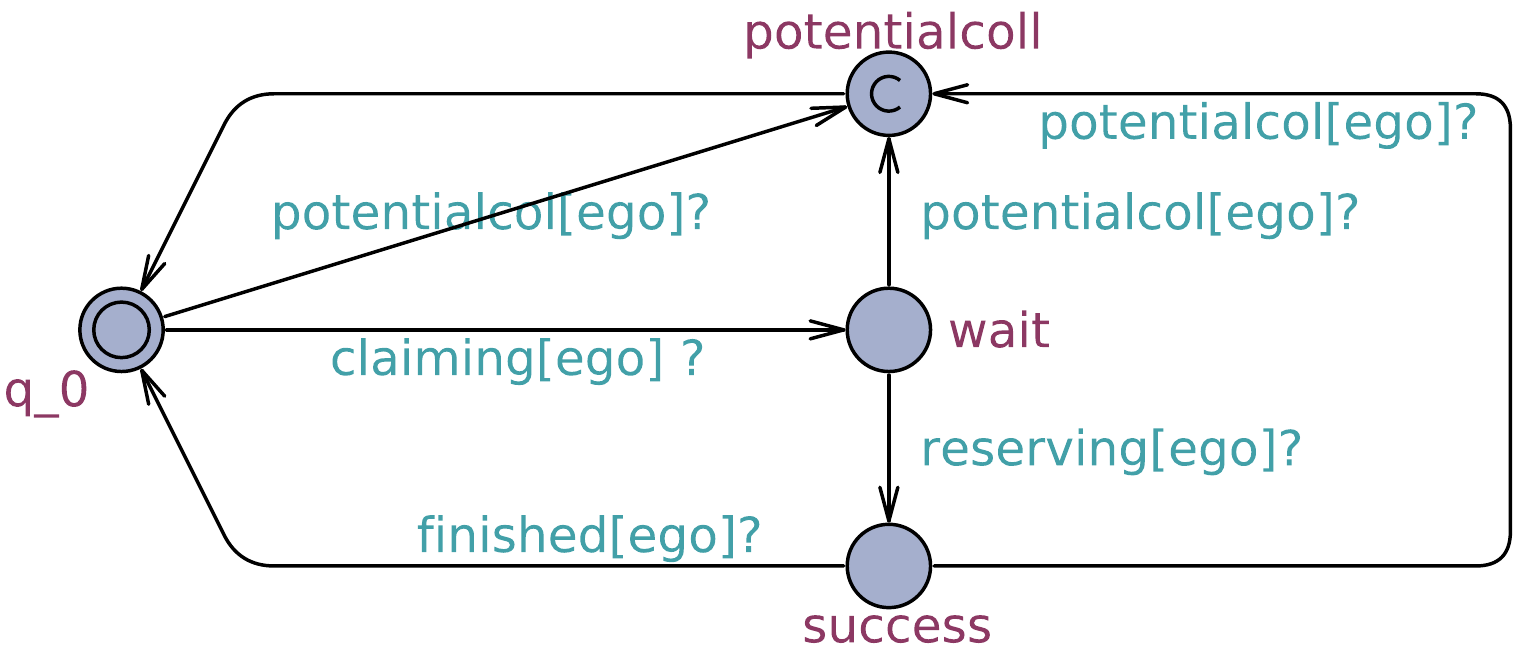}
	\caption[]{\texttt{Observer(i)} checks for every instance of the lane change controller \texttt{LCP(i)}, if whenever car i claims a lane, it finally changes lanes, or if a potential collision occurs.}
	\label{fig:upp:Observer}
\end{figure}

To analyse liveness, we introduce a second Observer automaton \texttt{Observer(i)}, as depicted in Fig.~\ref{fig:upp:Observer}. For every instance \texttt{LCP(i)} of the lane change controller, we require an automaton \texttt{Observer(i)} which synchronises with \texttt{LCP(i)} over communication channels. E.g.\ on claiming a lane for car $A$, \texttt{LCP(A)} sends over the channel \texttt{claiming[A]} with which \texttt{Observer(A)} synchronises, such that both controllers simultaneously change to a new state. Upon reserving a lane, \texttt{LCP(A)} sends over \texttt{reserving[A]} and the Observer changes to a state \texttt{success}. We check the query
\begin{align}\label{q3}
  \text{\texttt{A<> (Observer(A).success or Observer(B).success or Observer(E).success)},}
\end{align}
which states, that \emph{finally in every trace}, at least one of the controllers \texttt{LCP(i)} is successful in changing a lane. Remember, that we generally expect query~\eqref{q3} to be successfully verified, as in our model at least car $C$ should be able to finally change a lane in every possible trace.

\subsubsection{Adaption 1} Without a time invariant in state $q_1$ of the original controller from \cite{HLOR11} and without respective time guards on the outgoing edges, query \eqref{q3} \emph{does not hold}.

The reason is that there exists a trace, where cars $A$ and $B$ both infinitely often claim lane $1$ without any elapse of time and thus both circle between their respective states $q_0$, $q_1$ and $q_2$ in a \emph{livelock}. As no time elapses, \texttt{LCP(E)} has no possibility of executing any transition and thus starves. This problem is easily solvable by introducing the invariant $x <= t_w$ to state $q_1$ and placing the guard $x >= t_w$ on the outgoing edges of $q_1$, as done in the UPPAAL automaton depicted in Fig.~\ref{fig:upp:LCP}. With these adaptions, we successfully show query~\eqref{q3} in less than $0.5$ seconds with a memory usage peak of $40$KB.

\subsubsection{Adaption 2} The verification query \eqref{q3} is already a \emph{weak liveness} property, as it shows that in every simulation trace, at least one of the controllers finally changes lanes. We refine this property to 
\begin{align}\label{q4}
  \text{\texttt{A<> Observer(i).success},}
\end{align}
which states for an arbitrary car identifier $i$, that the related car finally changes lanes. When considering only the first adaption, as anticipated, this property only holds for \texttt{LCP(E)}. The reason is, that there still exists a trace, where cars $A$ and $B$ both unsuccessfully try to change to lane $1$ infinitely often and thus creating a potential collision infinitely often, preventing both controllers from ever transforming their claim into a reservation.

To solve this, we introduce an additional state \texttt{q\_wait}, in which the controller is forced to wait for a bounded non-deterministic time. For now, we delimit this waiting time in \texttt{q\_wait} by its invariant $x\leq4$ and the guard $x\geq1$ on its outgoing edge. With this, cars $A$ and $B$ do not permanently block each other from changing a lane and we verify both
\begin{align}\label{q5}
  \text{\texttt{A<> Observer(A).success}} &\hspace{1.3cm}\text{ and }\hspace{0cm} &\text{\texttt{A<> Observer(B).success}}
\end{align}
in each less than $2.7$ seconds with a memory usage of each less than $76$KB. 

\subsection{Summary and extendability of the current implementation}\label{sec:extendability}
With the traffic situation from Fig.~\ref{fig:model} and the corresponding implementation, as described in this section, we presented one very specific scenario, designed for the following purposes:
\begin{itemize}
    \item Showing the absence of collisions between any cars (i.e. proving safety \eqref{formula:safe}) and
    \item Identifying and analysing the existing livelocks (cf. location $q_1$) and
    \item Eliminating the livelocks and showing liveness of the new controller.
\end{itemize}
For this, the restrictions for the scenario, e.g. on $3$ cars and $4$ lanes were reasonable. However, we also tried different scenarios, with different numbers of lanes and cars. Our liveness and safety properties were not violated for any of the considered numbers of lanes and cars. Up to $16$ parallel lanes were considered without any problems. However, we observed the following run-time issues when adding cars.

While run-time seemes to increase only linear by about $50$ ms each time when we add one lane, it appears to increase exponentially when adding a car. This observation is not surprising, as adding only one car \texttt{i} means adding two timed automata and one clock variable to the system: One timed automaton \texttt{LCP(i)} with its clock $x$ and one observer automaton \texttt{Observer(i)}. Consider for example the model from Fig.~\ref{fig:model} with one additional car. Now for property \eqref{q5}, UPPAAL takes $1025$ seconds to verify the query instead of the previously observed $2.7$ seconds for the three car scenario. While $1025$ seconds for four cars is still acceptable, after including a fifth car, UPPAAL could not finish the verification of query \eqref{q5} within one day.

Thus, for future considerations of our implementation where more than four cars should be considered, we would have to optimise our implementation first.
\section{Conclusion}\label{sec:conclusion}
We strengthened the MLSL{} approach from the group of Olderog \cite{HLOR11, HLO13, HS16}, by implementing their lane change controller for highway traffic in UPPAAL and successfully verified their safety property. We additionally optimised their controller by examining and implementing liveness properties into it.

\subsubsection*{Related Work}
There exist several approaches for analysis and control of traffic using intelligent transportation systems, where e.g.\ in \cite{LP11} traffic lights are used as a central control mechanism at intersections. The authors verify safety of their hybrid systems with the tool KeYmaera. There also exists an approach to synthesise intelligent traffic light control mechanisms with the UPPAAL extension Stratego \cite{E17}. The key idea of this approach is to minimise waiting times and energy waste.

Also various approaches for safe and autonomously driving systems were implemented during the DARPA Grand Challenge, where e.g. finite state machines were used to describe the autonomous behaviour of the cars \cite{OSR07, WGJG08}.

For a hazard warning extension of MLSL, a dedicated \emph{hazard warning controller} was implemented in UPPAAL \cite{OS17}. However, the hazard warning controller was focused on a timely warning message delivery via broadcast channels and did not use MLSL formulas. A combined proof of UPPAAL verification queries with a formal proof by induction was used to prove the timely warning delivery.

\subsubsection*{Future Work}
In the end of Sect.~\ref{sec:liveness}, we observe that cars $A$ and $B$ block each other on lane $1$ and suggest an adaption ensuring the liveness of the controllers. However, this adaption does not guarantee \emph{fairness}, as one of the cars could get the right of changing lanes arbitrarily more often than the other car. To overcome this problem, we could implement a notion of fairness into \texttt{LCP(i)}, where either car $A$ or car $B$ lets the other car go first, when they already got the right of way often enough. Also, we could use the UPPAAL extension for stochastical model checking (UPPAAL SMC) \cite{BDG12}, to analyse the probabilities of unfair behaviour. We could add prices to the transitions of our controller, which increase, when a car unsuccessfully claims too often.

In this paper, we only considered the lane change controller for highway traffic \cite{HLOR11}. An implementation of their lane change controller for country-roads \cite{HLO13} and the crossing controller for intersections \cite{HS16} would be highly interesting. Also, they published results on a relaxation of their assumption of \emph{perfect knowledge}, where the controllers communicate, to cope for the missing information. Also, for future considerations with more cars or different controllers, an optimisation of our implementation is of high interest, as described in Sect.~\ref{sec:extendability}.

Last but not least, for now we have the assumption of a constant speed. To verify properties in a more realistic scenario, our cars should be able to dynamically change their speed. To this end, we plan to implement an adaption of the existing UPPAAL Stratego distance controller from \cite{LMH15}, as described in Sect.~\ref{sec:uppaal-model}. Their adaptive cruise control implementation also minimises the distance between the $ego$ car and the car in front, which we could use to optimise the traffic flow in our scenario. With this, we could even extend our MLSL scenario to a platooning scenario (cf. PATH Project \cite{LGS98} and the European SARTRE project \cite{SAR12}). However, as outlined in Sect.~\ref{sec:uppaal-model}, the adaption of the distance controller from \cite{LMH15} poses some non-trivial challenges.

\noindent
\par\bigskip
\textit{Acknowledgements.} I would like to thank Marius Miku{\v{c}}ionis for his help with starting the UPPAAL implementation.

\bibliographystyle{eptcs}
\bibliography{bib}

\end{document}